\documentclass[preprint,preprintnumbers,amsmath,amssymb,nofootinbib,longbibliography]{revtex4-1}

\usepackage{amsfonts}
\usepackage{amssymb}
\usepackage{latexsym}
\usepackage{graphicx}
\usepackage[utf8x]{inputenc}
\usepackage{longtable}
\usepackage{dcolumn}
\usepackage{amsmath}
\usepackage{bm}
\usepackage{young}
\usepackage{color}

\usepackage[active]{srcltx} 

\usepackage{rotating}

\newcommand{\al}{\alpha}

\newcommand{\be}{\beta}

\newcommand{\si}{\sigma}
\newcommand{\Si}{\Sigma}

\newcommand{\De}{\Delta}

\newcommand{\rar}{\rightarrow}

\begin{document}

\title{The  He$_2^+$ molecular ion and the He$^-$ atomic ion in strong magnetic fields}

\author{J. C. Lopez Vieyra}

\email{vieyra@nucleares.unam.mx}

\author{A.V.~Turbiner}

\email{turbiner@nucleares.unam.mx}

\affiliation{Instituto de Ciencias Nucleares, Universidad Nacional
Aut\'onoma de M\'exico, Apartado Postal 70-543, 04510 M\'exico,
D.F., Mexico}

\begin{abstract}
We study the question of  existence, {\it i.e.} stability with respect to dissociation of the spin-quartet, permutation- and reflection-symmetric ${}^4(-3)^+_g$ ($S_z=-3/2, M=-3$) state of the $(\al\al e e e)$ Coulomb system: the ${\rm He}_2^+$ molecular ion, placed in a magnetic field $0 \le B \le 10000$\,a.u. We assume that the $\al$-particles are infinitely massive (Born-Oppenheimer approximation of zero order) and adopt the parallel configuration, when the molecular axis and the magnetic field direction coincide,
as the optimal configuration.
The study of the stability is performed variationally with a physically adequate trial function.
To achieve this goal, we explore several Helium-contained compounds in strong magnetic fields,
in particular, we study the spin-quartet ground state of the ${\rm He}^-$  ion,  and  the ground (spin-triplet) state of the Helium atom, both for a magnetic field in $100 \leq B\leq 10000$\, a.u.
The main result is that the ${\rm He}_2^+$ molecular ion in the state ${}^4(-3)^+_g$ is stable towards
{\it all} possible decay modes for magnetic fields $B \gtrsim 120\,$a.u.\ and with the magnetic field increase the ion becomes more tightly bound and compact with a cigar-type form of electronic cloud.
At $B=1000\,$a.u., the dissociation energy  of  ${\rm He}_2^+$ into ${\rm He}^- + \al$ is { $\sim 702\,$eV} and the  dissociation energy for the decay channel to ${\rm He} + \al + e $ is { $\sim  729\,$eV}, latter both energies are in the energy window for one of the observed absorption features
of the isolated neutron star 1E1207.4-5209.
\end{abstract}

\maketitle

\section{Introduction}

It seems obvious that the chemical composition of the  atmosphere of a magnetic white dwarf or a magnetized neutron star can not be established as long as we lack   reliable information about the behavior of
many-body Coulomb systems, especially  about simple molecules and atoms in the presence of strong magnetic fields. So far, only Hydrogen-like and Helium-like atomic systems, and H$_2^+$-type molecular systems have been studied to a certain depth. There are indications that in addition to traditional atoms and molecules other non-standard, exotic atomic and molecular systems can also exist in strong magnetic fields (see for example, \cite{Turbiner2006309,PhysRevA.81.042503}). In a recent discovery \cite{2013PRL111}   it was found that even the ${\rm He}^-$ atomic ion becomes stable for magnetic fields $B \gtrsim 0.13\,$a.u.
with the spin doublet state ${}^2(-1)^+$ as ground state at first and then, for $B\gtrsim
0.74\,$a.u. all electron spins get aligned and antiparallel to a magnetic field direction: the
corresponding state ${}^4(-3)^+$ becomes the ground state of the system (see also \cite{PhysRevA.89.052501}). It happens in spite of the fact that Helium belongs to the most inert (closed shell) atomic systems. 
Needless to say that Helium has a rich chemistry even in the absence of intense magnetic fields (see e.g. \cite{Grandinetti2004IJMSp}).

Usually, investigations of the Coulomb systems in strong magnetic fields (unreachable in the lab) are justified by the fact there exists a strong magnetic field on the surface of many neutron stars $B\sim 10^{11-13}\,$G and of some highly-magnetized white dwarfs $B\sim 10^{8-9}\,$G, see e.g. review \cite{GarciaBerro2016}. In general, magnetic fields can reach $B\sim 10^{15}\,$G, or even higher, in the case of the so-called magnetars - the neutron stars with anomalously large surface magnetic field. While there is the evidence for the presence of Helium in the atmosphere of magnetic white dwarfs~\cite{JordanSchmelcher:1998}, there is no similar understanding about Helium in whatsoever form in atmospheres of neutron stars, and in general about their chemical content that can satisfactorily explain the observations.

The  discovery of absorption features at $\sim$0.7 KeV and $\sim$1.4 KeV in the X-ray spectrum of the isolated neutron star 1E1207.4-5209 by Chandra X-ray observatory~\cite{2002ApJ...574L..61S}, and its further confirmation by XMM-Newton X-ray observatory \cite{2002ApJ...581.1280M} motivated to perform studies of atoms and molecules in a strong magnetic field. At present there is a number of neutron stars whose atmospheres are characterized by absorption features: all of them are waiting to be solidly explained. These observations make clear that a detailed study of traditional atomic-molecular systems is needed, as well as for a search for new exotic chemical compounds which exist in a strong magnetic field only 
(see e.g. \cite{Turbiner2007,LopezVieyra2007,PhysRevA.75.053408,LopezVieyra2007b}).

As a result of such investigations,   a model of helium-hydrogenic molecular atmosphere of the neutron star 1E1207.4-5209 was proposed which is based on the assumption that the most abundant components in the atmosphere are the exotic molecular ions ${\rm He}_2^{3+}$ and ${\rm H}_3^{2+}$, with the presence of ${\rm He}^{+}, ({\rm HeH})^{2+}, {\rm H}_2^{+}$ subject to a surface magnetic field $\approx 4.4 \times 10^{13}$\,G (see \cite{Turbiner:2005m}). Conjectures about the absence of  hydrogen envelopes in some neutron stars have also motivated the study of atmospheres composed of neutral helium.  However, those simple models appear to be in conflict with observations. Models of atmosphere, composed with a large abundance of molecular systems containing helium, have been later suggested (see \cite{Turbiner:2005m}, \cite{Kerkwijk:2007} and references therein). For reasons which are not completely clear to the present authors it has been emphasized \cite{Kerkwijk:2007} that the ${\rm He}_2^+$ molecule has to play a particularly important role. This molecular system exists in a field-free case in the spin-doublet, nuclei-permutation-antisymmetric $(u)$,
reflection-symmetric ({\it with respect to any plane containing the internuclear axis}) $(+)$ ground state ${}^2{}0^+_u$ \cite{Pauling1933JChPh}, usually denoted as ${}^2{\Si_u}^+$. It is a rather compact system  characterized by a small dissociation energy $\sim 2.5$\, eV into ${\rm He}^+ ({}^2{\rm S}) + {\rm He} ({}^1{\rm S})$. The lowest spin-doublet, nuclei-permutation-symmetric $(g)$ excited state ${}^2 0^+_g$ \cite{Pauling1933JChPh}, usually denoted as ${}^2{\Si_g}^+$, is repulsive. It is essentially unbound with shallow van-der-Waals minimum at large internuclear distance, see \cite{Grandinetti2004IJMSp} and references therein. It took us a number of years   to perform a quantitative study of this particular system in the presence of a strong magnetic field, which is the subject of the present work. We are not aware of any similar previous study.

It is quite common in the field-free case that the ground state of simple atoms and molecules be characterized by the lowest possible total electron spin. Since the first qualitative studies of atomic and molecular systems
\cite{Kadomtsev:1970,Kadomtsev:1971b,Kadomtsev:1971a,Kadomtsev:1972,Ruderman:1971,Ruderman:1974} it became clear that in sufficiently strong magnetic fields the ground state is eventually realized by a state where the spins of all electrons are antiparallel to the magnetic field direction. Thus, the total electron spin takes maximal value as well as its total projection. It implies that the ground state depends on the magnetic field strength: there exists one (or several) threshold magnetic field for which one type of ground state changes to another one. This phenomenon was quantitatively observed for the first time for the ${\rm H}_2$ molecule. It was shown that spin-singlet ground state ${}^1\Si_g$, for small magnetic field, changes for intermediate fields $B\gtrsim 0.2\,$a.u. to the unbound (repulsive) ${}^3\Si_u$ state as the ground state (precisely in the domain of magnetic fields typical of magnetic white dwarfs). While for stronger magnetic fields $B \gtrsim 12.3\,$a.u., the ground state of the hydrogen molecule is realized by spin-triplet state ${}^3\Pi_u$ (see~\cite{Detmer:1998a} and references therein). Another recent example, which was mentioned above, is the case of the ${\rm He}^-$ atomic ion where the ground state is realized first by the spin doublet state ${}^2(-1)^+$ for $B \gtrsim 0.13\,$a.u., and later by the spin-quartet state ${}^4(-3)^+$,  for magnetic fields $B\gtrsim 0.74\,$ a.u., where all electron spins are aligned antiparallel to a magnetic field direction. In general, the phenomenon of change of the ground state nature with a magnetic field strength in traditional atomic systems was known since a time ago, see e.g.~\cite{IvanovJPB34_2001}  and reference therein.

The aim of this article is to perform a variational study of the ${\rm He}_2^+$ molecular
ion subject to a strong magnetic field in the state ${}^4(-3)^+_g$ , when all electron spins get oriented
anti-parallel to the magnetic field direction, the electronic total angular momentum projection
is equal to $M=-3$
\footnote{To avoid a contradiction with the Pauli principle, thus, the appearance of the Pauli forces,
it is further assumed that all three electrons have different magnetic quantum numbers, i.e. $m_1=0,m_2=-1,m_3=-2$},
and to show that the system is {\it stable} towards all possible decays or dissociation. Further it is naturally assumed that ${}^4(-3)^+_g$ is the ground state. Due to extreme technical complexity we do not discuss other states of ${\rm He}_2^{+}$ and leave the question about evolution of the type of the ground state with magnetic field changes for a future publication.
Since our study is limited to the question of the existence and stability of this system in a certain state, the main attention is devoted to the exploration of all possible decay channels. A natural assumption about the optimal (equilibrium) configuration with minimal total energy is one which is achieved in the parallel configuration: where the internuclear axis connecting  the two massive $\alpha$-particles (${\rm He}$ nuclei) is situated along the magnetic line.

Another aim of the article is to continue to study the ${\rm He}^-$ atomic ion in spin-quartet state ${}^4(-3)^+$ for strong magnetic fields $B \leq 1$\,a.u., which was initiated in \cite{2013PRL111}. This study is necessary   due to the possible decay mode ${\rm He}_2^{+}$ into ${\rm He}^- + \al$.

The consideration is non-relativistic, based on a variational solution of the Schr\"odinger equation. The magnetic field strength is restricted to magnetic fields { $B \leq 10000$\,a.u. ($=2.35\times 10^{13}\,$G)  below the relativistic Schwinger limit}. Also it is based on the Born-Oppenheimer approximation of zero-th order: the particles of positive charge ($\alpha$-particles) are assumed to be infinitely massive.

Atomic units are used throughout ($\hbar = m_e = e = 1$). The magnetic field $B$ is given in a.u., with $B_0 = 2.35 \times 10^9$ G. For energies given in eV,  the conversion $1\,{\rm a.u.} = 27.2\,{\rm eV}$ was used.
All energies, which are mentioned in the article, are the total energies (with spin terms included) if it is not indicated otherwise.

\section{${\rm He}_2^+$ Hamiltonian}

\begin{figure}
 \includegraphics[angle=0,width=100mm]{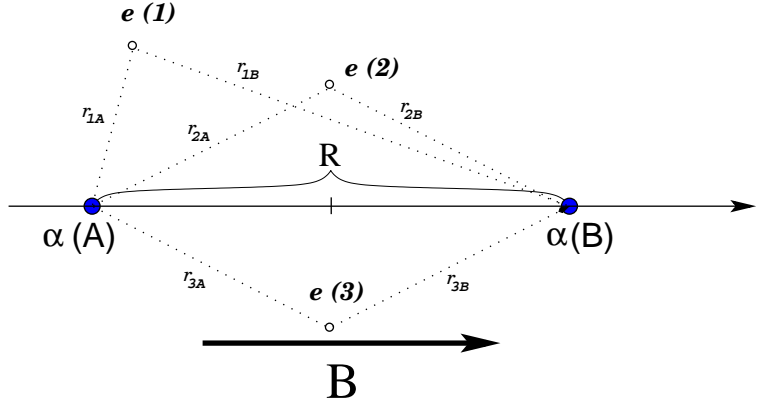}
 \caption{\label{fig1}Geometrical setting and notations for the ${\rm He}_2^+$ molecular ion in the
presence of a magnetic field $\mathbf{B}$ aligned parallel to the molecular axis.}
\end{figure}

The non-relativistic Hamiltonian for a three-electron diatomic molecule with fixed nuclei
$\sf A,B$ in uniform constant magnetic field $ \bm{B}=B  \bm{e}_z$, directed along $z$-axis, with vector potential in the symmetric gauge ${\bf A}=\frac{1}{2}{\mathbf B} \times {\mathbf r}$
is given by
\begin{equation}
\label{H}
  {\cal H}\ = - \sum_{i=1}^3 \left( \frac{1}{2} {\nabla}_{i}^2\ +
  \ \sum_{\eta= \sf A,B} \frac{Z_\eta}{r_{i\eta}} \right)
  \ + \ \sum_{{i=1}}^3 \sum_{{j>i}}^3 \ \frac{1}{r_{ij}}\ +\frac{B^2}{8} \sum_{i=1}^{3} \
{\rho_i}^2
+ \frac{B}{2} (\hat L_z +2\hat S_z )  + \frac{Z_{\sf A} Z_{\sf B}}{R}\, ,
\end{equation}
where ${\nabla}_{i}$ is the 3-vector
of the momentum of the $i$th electron, $Z_\eta$ is the  charge of the nucleus $\eta= \sf A,B$, the
terms  $-Z_\eta/r_{i\,\eta},$ correspond to the Coulomb interactions of the electrons with each
charged nuclei ($r_{i\,\eta}=|\bm{r}_{i\,\eta}|$ is the distance
between the $i$-th electron and the $\eta$-nuclei), the three terms  $1/r_{ij}$ ({\small
$j>i=1\ldots 3$}) are the
inter-electron Coulomb repulsive interactions ($r_{i\,j}$ are the distances between the $i,j$-th electrons), and the term $+Z_{\sf A} Z_{\sf B} /R$ is the
classical Coulomb repulsion energy between the nuclei, where $R$ is the internuclear distance (see Fig. (\ref{fig1}) for notations).
The Hamiltonian (\ref{H}) includes, the paramagnetic terms  $\frac{1}{2}\
\bm{B}\cdot\bm{l_i}$ as well as the spin Zeeman-term $\bm{B}\cdot\bm{s}_i$ for the interaction of the magnetic field with the spin, and  the diamagnetic term  $ \frac{\bm{B}^2}{8}{\rho_i}^2$,
with ${\rho_i}^2=x_i^2+y_i^2$ for each electron, $i=1,2,3$.   If the magnetic field is directed
along the  $z$~direction and parallel to the internuclear axis, the component of the total angular momentum along the $z$-axis $M$, the total spin $S$, the $z$~projection of the
total spin $S_z$, and the total $z$~parity $\Pi_z$ are conserved quantities. Sometimes, during the text the spectroscopic notation $\nu^{2S+1}M^{\Pi_z}$  (with standard labels $\Si, \Pi, \De \ldots$ for $|M|=0,1,2$ etc) is used for the electronic states. Here $\nu$ stands for the degree of excitation for given (fixed) symmetry. In our case of a homonuclear diatomic molecule an additional subscript $g/u$ (gerade/ungerade) indicates a symmetric/antisymmetric state with respect to the permutation of the identical nuclei.

\section{Ground state in a strong magnetic field (generalities)}

Before approaching to concrete calculations a description of the ground state of a Coulomb system
of $k$ electrons and several heavy charged centers in a strong magnetic field should be given.
In fact, a complete qualitative picture was presented in the pioneering works by Kadomtsev-Kudryavtsev~
\cite{Kadomtsev:1970,Kadomtsev:1971a,Kadomtsev:1971b,Kadomtsev:1972} and Ruderman \cite{Ruderman:1971,Ruderman:1974}:

{\it Observation.}

\begin{quote}

  (i) All spins of electrons are oriented antiparallel to the magnetic field direction. Hence, the total electronic spin
  projection is $-\frac{k}{2}$ ,

  (ii) All heavy centers are situated on a magnetic line. Hence, there is no gyration,

  (iii) Electronic magnetic quantum numbers are different and take values \linebreak
  $0, -1, -2, \ldots ,~{-(k-1)}$, hence, the total magnetic quantum number of the system $M=-\frac{k(k-1)}{2}$. This configuration does not contradict to the Pauli principle and implies vanishing Pauli forces.

\end{quote}

We are not familiar with a rigorous proof of the validity of this observation in general.

For Hydrogen atom, $k=1$, validity of this observation, see item (iii), was explicitly checked by Ruder et al \cite{Ruder:1994} and for other one-electron systems in \cite{Turbiner2006309} - it was shown the ground state corresponds to $M=0$. For Helium atom it was checked in \cite{Becken:2000} - the ground state was 
$(1s_0 2p_{-1})$ type, thus, $M=-1$ and for Lithium atom it was $(1s_0 2p_{-1} 3d_{-2})$ type, thus, $M=-3$ \cite{PhysRevA.70.033411} as well as for ${\rm He}^-$ \cite{2013PRL111}. For two-electron molecules ${\rm H}_2, {\rm HeH}^+, {\rm He}_2^{++}$ it was checked that lowest energy occurs at $M=-1$ comparing to $M=0,-2$, see \cite{Detmer:1998a},\cite{2007JPhB...40.3249T},\cite{PhysRevA.81.042503}, respectively.

\section{Trial Functions (general)}

The variational method is used to study the   state ${}^4(-3)^+_g$   of the ${\rm He}_2^+$ molecular
ion  (with infinitely massive centers) placed in a uniform magnetic field parallel to the molecular
$z$-axis. In general, the wavefunction of the electronic Hamiltonian (\ref{H})  with two identical
nuclei can be written in the form
\begin{equation}
\label{ansatz}
 \psi({\mathbf r}_1,{\mathbf r}_2, {\mathbf r}_3)  = (1+\sigma_N\,P_{\sf AB}){\cal A} \left[ \,
\phi({\mathbf r}_1,{\mathbf r}_2, {\mathbf r}_3) \chi \, \right]\,,
\end{equation}
where $\chi$ is a three-electron spin eigenfunction corresponding either to a total electronic spin
$S=1/2$ or $S=3/2$. Here ${\mathbf r}_{1,2,3}$ are position vectors of the first, second, third electrons,
respectively, see Fig.\ref{fig1}.
The function $\phi({\mathbf r}_1,{\mathbf r}_2, {\mathbf r}_3) $
is a three particle  orbital function, $P_{\sf AB}$ is the permutation operator of the two identical nuclei ($\sigma_N=\pm 1$ for {\it gerade/ungerade} states, respectively). The operator ${\cal A}$ is the three-particle antisymmetrizer:
\begin{equation}\label{Asym}
 {\cal A} = 1 - P_{12} - P_{13} - P_{23} + P_{231}  + P_{312}\,.
\end{equation}
Here, $P_{ij}$ represents the permutation $i \leftrightarrow j$, and
$P_{ijk}$ stands for the permutation of $123$ into $ijk$.
For strong magnetic fields which are typically present in the atmosphere of neutron stars,
a natural expectation is that the ground state corresponds to the case when all electron spins are aligned antiparallel to the magnetic field, {\it i.e.}  $S_z = -3/2$ and, thus, with the three-electron spin eigenfunction $\chi$ being totally symmetric.
In this case, when $S_z = -3/2$, the trial function is written as
\begin{equation}
\label{ansatznospin}
 \psi^{ S_z=-3/2} ({\mathbf r}_1,{\mathbf r}_2, {\mathbf r}_3) = (1+\sigma_N\,P_{\sf AB}){\cal A} \left[ \,
\phi({\mathbf r}_1,{\mathbf r}_2, {\mathbf r}_3)  \, \right]  \beta(1)\beta(2)\beta(3)\,,
\end{equation}
with $ \phi({\mathbf r}_1,{\mathbf r}_2, {\mathbf r}_3) $ a properly chosen orbital function, which then antisymmetrized by ${\cal A}$. Here $\be(i),\ i=1,2,3$ is spin down function of $i$th electron (see below).

{
On the other hand, for states of the total spin projection $S_z=-1/2$ ($S=1/2$) we have two linearly independent spin eigenfunctions:
\begin{align}
 \label{chi12}
 \chi^{}_{1}&=\frac{1}{\sqrt{2}}[\beta(1)\alpha(2)\beta(3)-\alpha(1)\beta(2)\beta(3)] \ , \\
 \chi^{}_{2}&=\frac{1}{\sqrt{6}}[2\beta(1)\beta(2)\alpha(3)-\beta(1)\alpha(2)\beta(3)
 -\alpha(1)\beta(2)\beta(3)]\ ,
\end{align}
where $\al(i) (\be(i)),\ i=1,2,3$ are spin up (spin down) functions of the $i$th electron.
So, the general form of a spin projection $S_z=-1/2$  function has the form
\begin{equation}
 \label{Psiproduct}
 \Phi=\phi_1({\mathbf r}_1,{\mathbf r}_2, {\mathbf r}_3) \chi_{1}+\phi_2({\mathbf r}_1,{\mathbf r}_2, {\mathbf r}_3) \chi_{2}\,,
\end{equation}
where $\phi_1({\mathbf r}_1,{\mathbf r}_2, {\mathbf r}_3) $ and $\phi_2({\mathbf r}_1,{\mathbf r}_2, {\mathbf r}_3) $ are orbital functions. In particular we can choose these functions to be proportional and write
\begin{equation}
 \label{Psis12}
 \Phi=\phi({\mathbf r}_1,{\mathbf r}_2, {\mathbf r}_3) ( \chi_{1} + a \chi_{2})\,,
\end{equation}
where $a$ is a variational parameter\footnote{A similar treatment was used for the study of the Li atom in a magnetic field in \cite{QUA:QUA22217}.}. Thus, the trial function for states of spin $S_z=-1/2$ can be written as ({\sl cf.} (\ref{ansatz}))
\begin{equation}
\label{ansatzs12}
 \psi^{S_z=-1/2} ({\mathbf r}_1,{\mathbf r}_2, {\mathbf r}_3) = (1+\sigma_N\,P_{\sf AB}){\cal A} \left[ \,
\phi({\mathbf r}_1,{\mathbf r}_2, {\mathbf r}_3) ( \chi_{1} + a \chi_{2}) \, \right]\,,
\end{equation}
with $ \phi({\mathbf r}_1,{\mathbf r}_2, {\mathbf r}_3) $ a properly chosen orbital function.
}

\section{Trial Functions (coordinate parts)}

\label{sectiontrialf}

Variational trial functions are designed  following physical relevance arguments.  In particular, we construct wavefunction which allow us to reproduce both the Coulomb singularities of the potential and the correct asymptotic behavior at large distances (see, e.g. \cite{Turbiner:1984}).  Following such criterion we propose the function
\begin{equation}
\label{phi}
  \phi({\mathbf r}_1,{\mathbf r}_2, {\mathbf r}_3)  = \prod_{k=1}^{3}
                    \left(\rho_k^{|m_k|}  e^{ i{m_k} \phi_k}    e^{-\al_{k, {\sf A}} r_{k{\sf A}}-\al_{k, {\sf B}}r_{k{\sf B}} }  e^{- \frac{B}{4} \beta_k \rho_k^2} \right)
                    e^{\al_{12} r_{12} + \al_{13} r_{13} + \al_{23} r_{23}} \ ,
\end{equation}
which is a type of product of Guillemin-Zener type molecular orbital functions multiplied by the product of Landau type orbitals for an electron in a magnetic field. Here $\al_{k_{\sf A}}, \alpha_{k_{\sf B}}$  and $\beta_k$,\ $k=1,2,3$ are parameters. { In the case of the fully polarized state $S=3/2$ and in order to} avoid a contradiction with the Pauli principle, it is further assumed that all electrons have different magnetic quantum numbers in a certain minimal way: $m_1=0,m_2=-1,m_3=-2$,  hence, the total electronic quantum number is $M=-3$. It was already discussed in \cite{PhysRevA.81.042503} that this assumption seems obviously correct in the case of atoms and atomic ions, where the electrons are sufficiently close to each other, but not that obvious for the case of molecules for which the electrons are spread in space. All of them (or, at least, some of them) can be in
the same quantum state, with the same spin projection and
magnetic quantum number. This situation was observed for ${\rm H}_2$ and ${\rm H}_3^+$,
where in a domain of large magnetic fields the ground state was given by the state of maximal
total spin but with the electrons having the same zero magnetic quantum number (see
\cite{PhysRevA.81.042503} and references therein). However, for very strong fields the state of minimal total energy always corresponds to the situation described above with all electrons having different magnetic quantum numbers.

In (\ref{phi}) the variational parameters $\al_{k, {\sf A}}, \al_{k, {\sf B}}$ ($k=1,2,3$) have the meaning of screening (or anti-screening) factors (charges) for the nucleus ${\sf A,B}$ respectively, as it is seen from the $k$-th electron. The variational parameters $\beta_k$ are screening (or anti-screening) factors for the magnetic field seen from  $k$-th electron, and the  parameters   $\al_{ij}, i<j = 1\ldots 3$ ``measure" the screening (or anti-screening) of the inter-electron interaction. In total, the trial function (\ref{phi}) has 12 variational parameters in addition to the internuclear distance $R$ which can also be considered as a variational parameter.

The calculation of the variational energy  using the trial function (\ref{ansatznospin})-(\ref{phi}) involves two major parts:  (i) 9-dimensional numerical integrations which were implemented by an adaptive multidimensional integration routine ({\sl cubature})\cite{GenzMalik1980}, and (ii) a minimizer which  was implemented with the
minimization package TMinuit from CERN-lib (an old version of TMinuitMinimizer in the ROOT
system\cite{Root1997}, which allows fixing/releasing parameters, was recovered and adapted  to our purposes). The 9-dimensional integrations were carried out using a dynamical partitioning procedure: the domain of integration is manually divided into sub-domains following the profile of the integrand. Then each sub-domain is integrated using the routine {\sl CUBATURE}. In total, we have a subdivision to $\sim 2000$ subregions for the numerator and $\sim 2000$ for the denominator in the variational energy. With a maximal number of sampling points $\sim 2\times 10^8$ for the numerical integrations (it guarantees the relative accuracy $\sim 10^{-3} - 10^{-4}$ in integration) for each subregion, the time needed for one evaluation of the variational energy takes $5 \times 10^4\,$ seconds with 96 processors. It was checked that this procedure stabilizes the estimated accuracy to be reliable in the first two decimal digits. However, in order to localize a domain, where minimal parameters are, the minimization procedure with much less number of sample
points was used in each sub-domain and a single evaluation of the energy usually took $\sim 15-20\,$mins. Once a domain is roughly localized the number of sample points increased by factor $\sim 10^2$.
{ Final evaluation was made with $2\times 10^8$ sampling points and for the strongest fields $B=100,1000,10000$ a.u.  it was even $5\times 10^8$ with a subdivision of 7 subintervals in each $z$-domain}. Typically, a minimization procedure required several hundreds of evaluations. Computations were performed in parallel with a cluster {\it Karen} with 96 Intel Xeon processors at $\sim 2.70$GHz.

\section{Decay Channels, Dissociation}
\label{sectiondecays}

In this section we analyze different decay channels of the $S_z=-3/2$,  $M=-3$ state ${}^4(-3)^+_g$ of the ${\rm He}_2^+$ molecular ion in a magnetic field  in the range
{ $1\,{\rm a.u.} \leq B \leq 10000\,$a.u.} Possible decay channels that we consider for
the system $(\al \al e e e)$ placed in a magnetic field are
\[
\begin{array}{llr}
{\rm He}_2^+ & \to {\rm He}     + {\rm He}^+        & (a) \\
                         & \to {\rm He}^-  + {\alpha}             & (b) \\
                         & \to {\rm He} + \alpha + e               & (c) \\
                         & \to {\rm He}^+  + {\rm He}^+ + e & (d) \\
                         & \to {\rm He}_2^{2+}   +  e              & (e) \\
                         & \to {\rm He}_2^{3+}  +  e + e         & (f) \\
\end{array}
\]
A few remarks emerge immediately: in the  cases where there are free electrons
\footnote{For the total energy of a free electron $E_e$ (excluding the spin contribution)
in the magnetic field in the symmetric gauge the $z$-component of the angular momentum $L_z$
is conserved and the electron Landau levels are $ E_e = \hbar\omega_B(n + 1/2)$, where $n=n_\rho +
\frac{|m|+m}{2}=0,1,2\ldots$ All $m \leq 0$ states are degenerate. Here $\omega_B=\frac{eB}{ m_e c}$ is the cyclotron frequency.}
in the decay channel (channels (c)-(f)) we can  assume that the electrons are in the ground state $n=0,s_z=-1/2,m<0$ which yields $E=0$ (regardless of the magnetic quantum number $m$ carried by the electron, where the spin contribution is included). In Born-Oppenheimer approximation the energy of a free $\alpha$-particle in a magnetic field is zero by assumption. Also, it is known that the helium atom exists for
any magnetic field strength.  For magnetic fields $0\leq B \lesssim 0.75\,$a.u. the spin-singlet state $1{}^1{0}^+$ is the ground state. For $B \gtrsim 0.75\,$ a.u., the spin-triplet state (with $m=-1$)  $1{}^3(-1)^+$ becomes the ground state. For the ${\rm He}$ atom { in a magnetic field $B\leq 100$ a.u.,} the corresponding total energies collected in Table~(\ref{TableEs}) were taken from references~\cite{Becken:1999,Becken:2000}. Such energies were calculated for the infinite nuclear mass approximation and  include the spin contributions $B \, s_z$ for each electron. { For magnetic fields $B=1000$ and $10000$ a.u. a simple variational Ansatz with 5 variational parameters was used to estimate the value of the energy of the spin-triplet state of Helium (see section \ref{sectionHe} below for more details).}

To obtain the energy of the ${\rm He}^+$ ion we use  the basic result of Surmelian and O'Connel   for hydrogen-like atoms
\begin{equation}
\label{Hscaling}
 E(Z, B) = Z^2E(1, B/Z^2)\,,
\end{equation}
and use the data of  \cite{Kravchenko:1996} for the binding energies  of Hydrogen in a magnetic
field  (recalling  that instead of the total energy $E_T$,  in   \cite{Kravchenko:1996} the authors
reported the energies $E_b = (1+ m + |m|) \gamma/2  -E$, which coincide with the binding energies
$\epsilon =  \gamma/2 - E$  for $m<0$).   The Zeeman contribution to the total energy $B\, s_z$
due to the spin of the electron is not taken into account in the results appearing in
\cite{Kravchenko:1996}. Such contribution  for the case of a spin antiparallel to the magnetic
field ($s_z=-\frac{1}{2}$) is $E_{\rm spin}= - B/2\,$a.u. and was added to the results collected
in Table~\ref{TableEs} to obtain the total energies of ${\rm He}^+$.

The molecular system $(\al \al e)$ (${\rm He}_2^{3+}$ molecular ion): accurate variational calculations in equilibrium configuration parallel to the magnetic field for the ground
state $1\sigma_g$ were carried out in detail in \cite{Turbiner2006309, 2007IJMPA..22.1605T} for the range of magnetic fields $100\, {\rm a.u.} \lesssim B \lesssim  B_{\rm Schwinger}$ where $B_{\rm Schwinger} = 4.414\times 10^{13}\,$G is the non-relativistic limit.  It was found that for magnetic fields $10^2 \lesssim B \lesssim 10^3\,$a.u.
the system ${\rm He}_2^{3+}$  is unstable towards  decay to ${\rm He}^+ + \alpha$.  Thus, in principle we can neglect in our considerations the decay channel (e) above. Nonetheless, at $B \gtrsim 10^4\,$ a.u., this compound becomes the system with the lowest total energy among the one-electron helium (helium-hydrogen) chains (for details see \cite{Turbiner2006309}).

The molecular system $(\alpha \alpha e e)$ ${\rm He}_2^{2+}$ molecular ion: this molecule was studied in detail in \cite{2006PhRvA..74f3419T} in the domain of magnetic fields $B = 0- B_{\rm Schwinger}$. It was shown that the lowest total energy state depends on the magnetic field
strength and  evolves from the spin-singlet ${}^1\Sigma_g$  metastable state at $0 \leq B \lesssim
0.85\,$ a.u. to a repulsive spin triplet ${}^3\Sigma_u$ (unbound state) for
$0.85\lesssim B \lesssim 1100\,$a.u. and then to a strongly bound triplet state  ${}^3\Pi_u$ state.
Hence, there exists quite a large domain of magnetic fields where the ${\rm He}_2^{2+}$ molecule is unbound and represented by two atomic helium ions in the same electron spin state but situated
at an infinite distance from each other.

\section{${\rm He}^-$ Revisited}
\label{sectionHeminus}

In order to have a complete understanding about the stability of the molecular ${\rm He}_2^+$ ion  in magnetic fields, we need to
extend the  study on the ${\rm He}^-$ atomic ion  (three electron atomic system $(\al,e,e,e)$)  in magnetic fields
to the regime of very strong fields $B\gg 1$a.u.
In this section we review the basic notions for the study of the ${\rm He}^-$ ion in   magnetic fields.
In  \cite{2013PRL111} it was found that the ground state of ${\rm He}^-$ in a magnetic field is realized by a spin-doublet
 $^2(-1)^{+}$ at $0.74\, {\rm a.u.} \gtrsim B \gtrsim 0.13\, {\rm a.u.}$ and it becomes a fully polarized spin-quartet $^4(-3)^{+}$ for larger magnetic fields.
Thus, we will extend that study of  the ${\rm He}^-$ ion in strong magnetic fields,  in the fully polarized, spin quartet $S=3/2$, state only.
For more details the reader is addressed to the reference   \cite{2013PRL111}.

The non-relativistic Hamiltonian for an atomic system of three-electron and  one infinitely massive center of charge $Z$    in a
magnetic field (directed along the $z$-axis and taken in the symmetric gauge)   is
\begin{equation}
\label{HHe-}
\begin{split}
  {\cal H}\ & =  - \sum_{k=1}^3 \left( \frac{1}{2} {\nabla}_{k}^2\ +
  \  \frac{Z}{r_{k}} \right) \ + \ \sum_{{k=1}}^3 \sum_{{j>k}}^3
  \ \frac{1}{r_{kj}}\  +\frac{B^2}{8} \sum_{k=1}^{3}
  \ {\rho_k}^2  + \frac{B}{2} (\hat L_z +2\hat S_z ) \, ,
\end{split}
\end{equation}
where ${\nabla}_{k}$ is the 3-vector momentum of the $k$th electron,
$r_{k}$ is the distance between the $k$th electron and the nucleus,
$\rho_k$ is the distance of the $k$th electron to the $z$-axis, and
$r_{kj}$ \,$(k,j=1,2,3)$ are the inter-electron distances. $\hat L_z$
and $\hat S_z$ are the $z$-components of the total angular momentum and
total spin operators, respectively. Both $\hat L_z$ and $\hat S_z$ are
integrals of motion and can be replaced in (\ref{H}) by their eigenvalues $M$ and $S_z$
respectively.  For He$^-$ the nuclear charge is $Z$=2.  The total
spin ${\hat S}$ and $z$-parity $\hat{\Pi}_z$ are also conserved
quantities.  The spectroscopic notation $\nu{}^{2S+1}M^{\Pi_z}$ is
used to mark the states, where $\Pi_z$ denotes  the $z$~parity eigenvalue
$(\pm)$, and  the quantum number $\nu$ labels the degree of excitation.
For states with the same symmetry, for the lowest energy states at $\nu=1$
the notation is ${}^{2S+1}M^{\Pi_z}$. We always consider states with $\nu=1$
and $S_z=-S$ assuming they correspond to the lowest total energy states of
a given symmetry in a magnetic field.

\subsection{Trial functions}

The spin $S=3/2$ state ${}^4(-3)^{+}$ of the system $(\al,e,e,e)$ in a magnetic field is
described by the trial function
\begin{equation}
\label{GStrialfunct}
 \psi(\vec{r}_1,\vec{r}_2,\vec{r}_3) = {\cal A} \left[ \,
   \phi(\vec{r}_1,\vec{r}_2,\vec{r}_3) \, \right]\,,
\end{equation}
where   ${\cal A}$ is the three-particle antisymmetrizer (\ref{Asym})  and  $\phi(\vec{r}_1,\vec{r}_2,\vec{r}_3)$ is the explicitly
correlated orbital  function
\begin{equation}
\label{He-phi}
  \phi( \vec{r}_1,\vec{r}_2,\vec{r}_3)  =  \left( \prod_{k=1}^{3}
                   \rho_k^{|M_k|}  e^{ i{M_k} \phi_k}    e^{-\al_{k} r_{k} - \frac{B}{4} \beta_k \rho_k^2 } \right)
                    e^{\al_{12} r_{12} + \al_{13} r_{13} + \al_{23} r_{23}} \ ,
\end{equation}
where $M_k$ is the   magnetic quantum number
and $\al_k$, $\beta_k$ and $\al_{kj}$ are non-linear variational
parameters for each electron $k=1,2,3$. In total, the trial function (\ref{GStrialfunct}) contains
9 variational parameters.  The function (\ref{GStrialfunct}) is a
properly anti-symmetrized product of $1s$ Slater type orbitals, the lowest
Landau orbitals and the exponential correlation factors $\sim\exp{(\al\, r_{kj})}$.

\bigskip

The spin 3/2 state ${}^4(-3)^{+}$ of the system $(\al,e,e,e)$ in a magnetic field is
described by the trial function (\ref{GStrialfunct}) with $M_1=0, M_2=-1, M_3=-2$.
Due to the spin Zeeman contribution, the energy of this (spin $S=3/2$)
state decreases rapidly and monotonically with the magnetic field increase  and becomes the  (stable)
ground state for $B\gtrsim 0.7$\,a.u.

In \cite{2013PRL111}, we made a study of the  $(\al,e,e,e)$ atomic system in  magnetic fields $B\leq 100\,$a.u.
Here we extend that study for magnetic fields up to \hbox{$B=10000\,$a.u.}
In particular, we improve the value of the total energy at $B=100\,$a.u. from $E_T = -13.29\,$a.u.
(as quoted in ref. \cite{2013PRL111}) to $E_T = -13.38\,$a.u., {\it i.e.} by $\sim 0.1\,$a.u.
(see Table \ref{TableEs}).

The variational method used to find the energy of the system with the trial function (\ref{GStrialfunct}) involves two major procedures
of numerical minimization and integration. This was already described  above for the case of the ${\rm He}_2^+$ molecular ion (see section \ref{sectiontrialf}).
In particular, for strong magnetic fields, a reliable minimization depends on the accuracy of the variational energies {\it i.e.} on the accuracy of the numerical
9-dimensional integrations.  Our strategy to find the minimal energy was first to make approaching minimizations with relatively low accuracy in the integrations and then followed by  a manual scanning of the energy dependence on each variational parameter   with high accuracy in the numerical integrations.   For our final results
we used a partition of the integration domain into 4800 subregions for the numerator and 4800 subregions for the denominator using 500 million points for each
numerical integration.  A single evaluation of the energy takes about 14 hrs of wall clock time using a cluster with 120 processors.

Our results for the spin 3/2 state ${}^4(-3)^{+}$ of the system $(\al,e,e,e)$ in  magnetic fields $B=100,1000, 10000$\, a.u. are collected in Table  \ref{TableEs} below.
These results indicate that as the magnetic field increases, the total energy of the ${\rm He}^-$ ion in the ${}^4(-3)^{+}$ state decreases, but at a slower rate in comparison to the total energy of the ${\rm He}_2^+$ molecular ion in the ${}^4(-3)^{+}$ state. Also, our results confirm that as the magnetic field increases, the  total energy of the
${\rm He}^-$ ion in the ${}^4(-3)^{+}$ state decreases more rapidly than the total energy of the ${\rm He}$ atom in the spin triplet state ${1}^3(-1)^+$ and, therefore, becoming more stable towards decay into ${\rm He}^-  \to  {\rm He} + \alpha$ (see below for our extended calculations of the ${\rm He}$ atom in the spin triplet state ${1}^3(-1)^+$).

\section{${\rm He}$ atom in strong magnetic fields revisited}
\label{sectionHe}

 The  assertion on the stability of the molecular ion ${\rm He}_2^+$ in strong magnetic fields requires a  full  understanding of the Helium atom and other Helium species in the presence of a strong magnetic field. It is known that the Helium atom exists for any magnetic field strength. Its ground state is realized by a singlet spin state $1{}^1S$
 at zero and small magnetic fields $0\leq B \leq 0.75\,$a.u. For larger magnetic fields the ground state is realized by the fully polarized spin triplet state $1{}^3(-1)^+$ \cite{Becken:2000}. Despite the fact that the Helium atom in magnetic fields is one of the most studied systems, all such studies, we are familiar with, are limited to magnetic fields up to $B=100\,$a.u. (see, e.g. \cite{Becken:2000}). For higher magnetic fields (up to the Schwinger limit $B_{\rm rel} = 4.414\times 10^{13}\,$G $\sim 18783\,$a.u.) the relativistic corrections to the energy seem to be relatively small.  Thus, we extend the study of the Helium atom up to the largest magnetic field $B=10^4\,$a.u. considered in the present study.  In particular, a full understanding of the  ${\rm He}$ atom in the spin triplet (ground) state ${1}^3(-1)^+$ at magnetic fields $10000\geq B\geq 100\,$a.u. is necessary.

\subsection{Hamiltonian}
The non-relativistic Hamiltonian  which described the Helium atom with an infinitely massive nucleus of charge $Z=2$ in a magnetic field oriented along the $z$-axis
is given  by
\begin{equation}
\label{HHe}
\begin{split}
  {\cal H}\ & =  - \sum_{k=1}^2 \left( \frac{1}{2} {\nabla}_{k}^2\ +
  \  \frac{Z}{r_{k}} \right) \ + \,  \frac{1}{r_{12}}\  +\frac{B^2}{8} \sum_{k=1}^{2}
  \ {\rho_k}^2  + \frac{B}{2} (\hat L_z +2\hat S_z ) \, ,
\end{split}
\end{equation}
where  the symmetric gauge ${\bf A}=\frac{1}{2}{\mathbf B} \times {\mathbf r}$ was used, and ${\nabla}_{k}$ is the 3-vector momentum of the $k$th electron,
$r_{k}$ is the distance between the $k$th electron and the nucleus,
$\rho_k$ is the distance of the $k$th electron to the $z$-axis, and
$r_{12}$   is the inter-electron distance. $\hat L_z$
and $\hat S_z$ are the $z$-components of the total angular momentum and
total spin operators, respectively. Both $\hat L_z$ and $\hat S_z$ are
integrals of motion and can be replaced in (\ref{HHe}) by their eigenvalues $M$ and $S_z$
respectively.   The total spin ${\hat S}$ and $z$-parity $\hat{\Pi}_z$ are also conserved
quantities.  The spectroscopic notation $\nu{}^{2S+1}M^{\Pi_z}$ is
used to mark the states, where $\Pi_z$ denotes  the $z$~parity eigenvalue
$(\pm)$, and  the quantum number $\nu$ labels the degree of excitation.
For states with the same symmetry, for the lowest energy states at $\nu=1$
the notation is ${}^{2S+1}M^{\Pi_z}$.

\subsection{Trial functions}
To study the ground state of the Hamiltonian (\ref{HHe}) we use the variational method with trial functions
chosen  according to the criterion of  physical relevance. The trial functions for the low lying states of (\ref{HHe}) can be written as
\begin{equation}
\label{psiHe}
\Psi  =  (1+\sigma_e P_{12} )
\big(
\rho_1^{|m_1|}  e^{ i{m_1} \phi_1} \rho_2^{|m_2|}  e^{ i{m_2} \phi_2}
  e^{-\al_{1} r_{1} - \al_{2 }r_{2} - \frac{B}{4} (\beta_1 \rho_1^2 + \beta_2 \rho_2^2 )  + \al_{12} r_{12} }
\big)\,,
\end{equation}
where $P_{12}$ is the permutation operator for the electrons ($1 \leftrightarrow 2$) and $\sigma_e = \pm 1$ corresponds to the
spin singlet ($\sigma_e=1$) and the spin triplet ($\sigma_e=-1$) eigenstates, and $m_{1,2}$ are the magnetic quantum numbers of electron ($1,2$ respectively).
The trial function (\ref{psiHe}) depends on 5 variational parameters   which account for  effective screened charges:
of the nucleus $\al_{1,2}$ (as seen by electrons 1,2), of the electrons moving in the magnetic field $\be_{1,2}$, and the effective charge of one of the electrons as seen from the other $\al_{12}$.

The ground state of the Helium atom at strong magnetic fields is realized by the spin triplet state $1{}^3(-1)^+$ corresponding to $\sigma_e=-1$ and  $m_1=0, m_2=-1$, ($M=m_1+m_2$) in (\ref{psiHe}). This simple Ansatz  gives an energy $E_T=-12.8215$ a.u. at $B=100$ a.u.,   which compared to the most accurate result $E_T=-13.1048$ a.u. in \cite{Becken:1999}, indicates that the relative difference provided by this trial function is $\sim 2\%$.
The total energy of the triplet state $1{}^3(-1)^+$ decreases as the magnetic field increases.
For a magnetic field $B=1000\,$a.u. our variational trial function   gives an energy of $E_T=-27.1738$ a.u.  while at  $B=10000\,$a.u. it  gives an energy of $E_T=-53.2011$ a.u.  It is worth to notice that the total energy of Helium in the spin triplet state $1{}^3(-1)^+$ lies higher than the total energy of the ${\rm He}^-$ ion
in the fully polarized state ${}^4(-3)^+$, and the energy difference increases with an increase of the magnetic field.

 The variational method used to find the energy of the Helium atom with the trial function (\ref{psiHe})
 involves two major procedures of numerical minimization (MINUIT) and multidimensional numerical integration (Cubature).
 Due to the axial symmetry of the problem the dimensionality reduces to five. The integrations are performed in double cylindrical
 coordinates $(z_1,\rho_1,z_2,\rho_2,\phi )$ (where $\phi$ is the relative azimuthal angle between the electrons). The manual partitioning
 includes five subdomains in each  $z$  coordinate,  three subdomains in each $\rho$  coordinate and one domain for $\phi$.   The maximal number of point used
 to evaluate the numerical  integrations is 50 millions.  Our results for this system are presented in Tables  \ref{TableEs} and \ref{heparams}  (see below).

\section{Results}

\subsection{Field free case: low lying states}

We have carried out variational calculations for the  field free ground state ${}^2\Sigma_u^+$ state  as well as  for the weakly bound excited state ${}^2\Sigma_g^+$ state of ${\rm He}_2^+$. The aim of this study is mainly to have an estimate of the accuracy of our variational calculations. Previous studies on one and two electron Coulomb systems in strong magnetic fields  have shown that simple trial functions of the type (\ref{phi}) which are built following the criterion of physical adequacy, have led to very accurate results for such systems (see \cite{Turbiner2006309,PhysRevA.81.042503}).

The results for the energy of the states ${}^2\Sigma_u$ and  ${}^2\Sigma_g$ using the trial function (\ref{ansatzs12})
are collected in the Table  \ref{tables12B0}. From such results, and comparing to the most accurate results to date for such states
\cite{doi:10.1063/1.3692800, Xie2005}, we can conclude  that the energies obtained with our 10-parameter trial function have a
relative accuracy  $\sim 1\%$.  It is worth to note that the level of accuracy provided by the trial function (\ref{ansatzs12}) is sufficiently high to observe the shallow minimum of the ${}^2\Sigma_g$ state, though the equilibrium distance seems to be slightly shifted in comparison to
the results of  \cite{Xie2005}.

In a magnetic field one can expect a relatively slow decrease in accuracy as the magnetic field increases.
A similar comparison for the energies of the Helium atom in a magnetic field $B\leq 100$\,a.u. obtained (a) with the two-electron 5-parameter trial function  (\ref{psiHe}) and  (b) with the more accurate energies  using a Gaussian basis set method with $\sim 4300$  two-particle
functions \cite{Becken:1999}, leads to the conclusion that even at $B=100$\,a.u. the relative accuracy is $\sim 2\%$.
Thus, we can estimate that our results for the ${\rm He}_2^+$ ion in magnetic fields $B\lesssim 100$\, a.u. have an accuracy of $\sim 2\%$ with a small decrease for higher magnetic fields. To confirm this conjecture, a separate study would be necessary.

\begin{table}
\begin{tabular}{|ccc|ccc|}
\hline
\multicolumn{3}{|c}{${}^2\Sigma_u$} &\multicolumn{3}{c|}{${}^2\Sigma_g$} \\ \hline
$E$ (a.u.)     &  $R_{eq}$ (a.u.)    &  &            $E$ (a.u.) & $R_{eq}$ (a.u.)  & \\
\hline
-4.955243      &  \ \quad 2.15  $(\star)$ &  &       -4.8653    & \quad 8.61 $(\star)$ & \\
-4.953765      &    2.042            &  &            -4.8651    &  8.742           & \\
-4.994644      &    2.042            & ${}^{(a)}$ &\ -4.9036    &  8.742      & ${}^{(b)}$\\
 \hline
\end{tabular}
\caption{\label{tables12B0}
    ${\rm He}_2^+$, field-free case:\\
    First row $\rar$ energy and equilibrium distance $(\star)$ for the ground state ${}^2\Si_u$ and excited state ${}^2\Si_g$ calculated with a trial function (\ref{ansatzs12}).\\
   Second row $\rar$ the energies calculated at the equilibrium distance taken from
   \cite{doi:10.1063/1.3692800, Xie2005}.\\
   Third row $\rar$ energy (rounded) and equilibrium distance for ${}^2\Si_u$ from
   \cite{doi:10.1063/1.3692800} ${}^{(a)}$ and for ${}^2\Si_g$ from \cite{Xie2005} ${}^{(b)}$.}
\end{table}

\subsection{$B=1\,$a.u.}

We begin our analysis for the
total  energy and equilibrium distance for the
 spin $S=-3/2$ state with $M=-3$ corresponding to the Hamiltonian (\ref{H})  for a magnetic field $B=1$\,a.u.

 For this magnetic field the total energy of the spin-quartet, $m=-3$  state   obtained with  the trial function (\ref{ansatznospin}-\ref{phi}),  is $E_T^{{\rm He}_2^+}({}^4(-3)^+ )=-4.02\,$ a.u. with an equilibrium distance $R_{\rm eq} = 2.18\,$ a.u. (see Table \ref{TableEs}).  The lowest energy state of ${\rm He}$
corresponding to the triplet state, and $m=-1$,  has a total energy  $E_T^{{\rm He}}(1{}^3(-1)^+) =-2.9655\,$~a.u., and  the total energy  of the ground state of  ${\rm He}^+$ ion,
is $E_T^{{\rm He}^+}(1s_0) = -2.4410 \,$~a.u.   From this considerations, it is clear that for
this magnetic field  the state ${}^4(-3)^+_g$ of ${\rm He}_2^+$ is unstable towards decay (channel (a)) to
\begin{equation}
\label{decaya}
{\rm He}_2^{+}({}^4(-3)^+_g) \to {\rm He}(1{}^3(-1)^+) +  {\rm He^+}(1s_0)\ ,
\end{equation}
since the total energy of the sub-products
$E_T^{{\rm He}}(1{}^3(-1)^+)   + E_T^{{\rm He}^+}(1s_0) =   -5.4065\,$ a.u.
is essentially lower than the total energy of the spin-quartet $m=-3$  state of ${\rm He}_2^+$.
It is also noteworthy to mention that the lowest state of ${\rm He}^+$ with  $|m|=2$, has a total
energy  of   $E_T^{{\rm He}^+}(3{\rm d}_{-2})=-0.7930\,$a.u. (including the contribution from the
spin Zeeman term) and, for this case,  the state
${}^4(-3)^+$ of ${\rm He}_2^+$ is stable towards the ($M$-conserved) decay into
\begin{equation}
{\rm He}_2^{+}({}^4(-3)^+_g) \nrightarrow {\rm He}(1{}^3(-1)^+) +  {\rm He^+}(3d_{-2})\,,
\end{equation}
since the total energy of the sub-products is
$E_T^{{\rm He}}(1{}^3(-1)^+_g)   + E_T^{{\rm He}^+}(3d_{-2}) =   -3.7585\,$a.u. Thus, ${\rm
He}_2^+$ in the state ${}^4(-3)^+_g$ is a metastable state.

For this magnetic field ${\rm He}_2^+({}^4(-3)^+_g)$ is also unstable towards decay into  two ${\rm
He}^+(1s_0)$ ions plus an electron infinitely separated (decay channel (d)). This separated system
has a total energy $-4.8820\,$a.u.
On the other side,  following the results summarized in Table (\ref{TableEs}) ${\rm He}_2^+({}^4(-3)^+_g)$ is stable towards decays into  ${\rm He}^-({}^4(-3)^+ ) +\alpha$ (channel (b)) or ${\rm He}(1{}^3(-1)^+ ) +\alpha + e$ (channel (c)). Decay channels (e) and (f) are not possible since the systems
${\rm He}_2^{2+}$ and ${\rm He}_2^{3+}$ either do not exist or are unstable.   For $B=1\,$a.u., the lowest energy state of the two electron molecular ion ${\rm He}_2^{2+}$ corresponds to a purely repulsive spin-triplet (unbound) state ${}^3\Sigma_u$ (for  $0.85\lesssim B \lesssim 1100\,$a.u. this system do not exist four-body bound state, it exists in a form of two separated helium ions He${}^+$ situated at an infinitely large distance from each other), and the one electron molecular ion ${\rm He}_2^{3+}$ in its ground  state ${1}\si_g$ is unstable towards decay into ${\rm He}^+ + \alpha$.

\subsection{$B=100\,$a.u.}

We continue our analysis for a magnetic field $B=100$\,a.u. For this magnetic field the total energy
of the quartet state  with $m=-3$  belonging to the Hamiltonian (\ref{H})   is
$E_T^{{\rm He}_2^+}({}^4(-3)^+_g )= { -22.46}\,$ a.u. with an equilibrium distance { $R_{\rm eq} = 0.432\,$}
a.u. (see Table \ref{TableEs}).

For this magnetic field the lowest energy state of ${\rm He}$ is the triplet state with magnetic quantum number $m=-1$,
and has a total energy   $E_T^{{\rm He}}(1{}^3(-1)^+) =  -13.1048\,$a.u., while the total energy of
the ground state of  the ${\rm He}^+$ ion is $E_T^{{\rm
He}^+}(1s_0) = -9.5605\,$ a.u. (including the spin Zeeman contribution).  From this we can
conclude that  the state  ${}^4(-3)^+_g$ is still unstable towards decay  (\ref{decaya}), since the
total energy of the sub-products is
$E_T^{{\rm He}}(1{}^3(-1)^+)   + E_T^{{\rm He}^+}(1s_0) = -22.6653\,$ a.u. which is slightly smaller than the total energy of
the  quartet $m=-3$  state of ${\rm He}_2^+$. However, it is clear form this comparison, that for some { $B > 100\,$a.u.}
the state ${}^4(-3)^+$ of ${\rm He}_2^+$ becomes stable towards decay into  ${\rm He} +  {\rm He^+}$ (see below).

Following the results summarized in Table (\ref{TableEs}) we conclude that ${\rm
He}_2^+({}^4(-3)^+_g)$ is stable towards decays into  ${\rm He}^-({}^4(-3)^+ ) +\alpha$ (channel
(b)) or ${\rm He}(1{}^3(-1)^+ ) +\alpha + e$ (channel (c)) or ${\rm He}^+(1s_0 ) +{\rm He}^+(1s_0 )  +
e$ (channel (d)) since the total energies of the corresponding separated subsystems in all these
channels lie above the total energy of the quartet state.  Decay channels (e) and (f) are also not
possible since, for this magnetic field,
the lowest energy state of the two electron molecular ion ${\rm He}_2^{2+}$ corresponds to a
purely repulsive triplet state ${}^3\Sigma_u$,  and the system ${\rm He}_2^{3+}$ in its ground
state ${1}\sigma_g$ is unstable towards decay into ${\rm He}^+ + \al$.

\subsection{$B=1000\,$a.u.}

Our variational result for the total energy
of the quartet state  with $m=-3$  belonging to the Hamiltonian (\ref{H})  for a magnetic field $B=1000$\,a.u.    is
$E_T^{{\rm He}_2^+}({}^4(-3)^+_g )= { -53.98}\,$ a.u. with an equilibrium distance $R_{\rm eq} = 0.196\,$ a.u. (see Table \ref{TableEs}).

Now, for the main decay channel (a)  the total energy of the sub-products  is $E_T^{{\rm He}}(1{}^3(-1)^+)   +
E_T^{{\rm He}^+}(1s_0) =   -47.4445 \,$a.u. which  lies higher
than the total energy of the quartet state of ${\rm He}_2^+$, and thus, the molecular ion ${\rm He}_2^+$
is  stable towards decay to ${\rm He}_2^+ \to {\rm He}     + {\rm He}^+ $.  The dissociation
energy for this channel is { $6.54\,$a.u. $=177.8\,$eV} at $B=1000\,$a.u.

For the case of channel (b), a direct comparison of the total energies of  ${\rm He}_2^+$ and  ${\rm He}^-$ at $B=1000\,$~a.u.
{ (for ${\rm He}^-$ we made an extension of the results in \cite{2013PRL111} at $B=1000$ a.u.)} indicates that ${\rm He}_2^+$ is also  stable towards decay to ${\rm He}_2^+  \nrightarrow {\rm
He}^-     +  \alpha $ { with a dissociation energy of $\sim 701.8$ eV.}  For the case of channel (c)     ${\rm He}_2^+  \to {\rm He} + \alpha + e$, the total
energy of the sub-products
of this decay is larger  than the total energy of the ${\rm He}_2^+ $ ion in the quartet state. So,
the system is also stable towards this decay channel { with a dissociation energy of $\sim 729.1$ eV.}  For the case of channel (d)
${\rm He}_2^+  \to {\rm He}^+  +  {\rm He}^+ + e$ the total energy of the
sub-products of this decay is larger  than the total energy of the ${\rm He}_2^+ $ ion in the
quartet state. { The dissociation energy in this case is $\sim 365.5$ eV.} For the case of channel (e)     ${\rm He}_2^+  \to {\rm He}_2^{2+} ({}^3\Pi_u) + e\,$, at $B=1000\,$a.u. the
total energy of the sub-products of this decay is larger
than the total energy of the ${\rm He}_2^+ $ ion in the quartet state. However, we should remember
that for this magnetic field  the ground state of the  two-electron
molecular ion ${\rm He}_2^{2+}$  is realized by a repulsive spin triplet state ${}^3\Sigma_u$ state
(for  $0.85\lesssim B \lesssim 1100\,$a.u. this system exists in the form of two
helium ions He${}^+$ situated at an infinitely large distance from each other { {\it i.e. } case (d)}), and  the  strongly bound
triplet state  ${}^3\Pi_u$ state becomes the ground state
at $B\gtrsim 1100\,$a.u. (see \cite{2006PhRvA..74f3419T}).

For the case of channel (f)     ${\rm He}_2^+  \to {\rm He}_2^{3+} (1\sigma_g) + 2 e\,$, at $B=1000\,$a.u. the
total energy of the sub-products of this decay is larger
than the total energy of the ${\rm He}_2^+ $ ion in the quartet state, and the corresponding dissociation energy is $\sim 934.2$ eV.  For this magnetic field the
ion ${\rm He}_2^{3+}$
is stable towards ${\rm He}_2^{3+} \to {\rm He}^{+} + \alpha$ and it is the most bound one-electron
system made from protons and/or $\alpha$ particles for $B>1000\,$a.u.

\subsection{$B=10000\,$a.u.}

Our variational result for the total energy
of the quartet state  with $m=-3$  belonging to the Hamiltonian (\ref{H})  for a magnetic field $B=10000$\,a.u.    is
$E_T^{{\rm He}_2^+}({}^4(-3)^+_g )= { -114.9}\,$ a.u. with an equilibrium distance $R_{\rm eq} = 0.098\,$ a.u. (see Table \ref{TableEs}).
At this magnetic field the numerical integrations with different accuracies (maximal number of points) indicate that the energy has a relative accuracy
of $0.1\,$a.u. This may result in differences of a few eV  in the transition energies.

For the main decay channel (a)  the total energy of the sub-products   is $E_T^{{\rm He}}(1{}^3(-1)^+)   +
E_T^{{\rm He}^+}(1s_0) =   - 92.7118 \,$a.u. which  lies higher
than the total energy of the quartet state of ${\rm He}_2^+$, and thus, the molecular ion ${\rm He}_2^+$
is  stable towards decay to ${\rm He}_2^+ \to {\rm He}     + {\rm He}^+ $.  The corresponding dissociation
energy for this channel is { $22.2\,$a.u. $=603.5\,$eV} at $B=10000\,$a.u.

For the case of channel (b), a direct comparison of the total energies of  ${\rm He}_2^+$ and  ${\rm He}^-$ at \hbox{$B=10000\,$~a.u.}
(in section \ref{sectionHeminus} we carried out an extension of the results in \cite{2013PRL111}   for ${\rm He}^-$ up to \hbox{$B=10000$ a.u.})   indicates that ${\rm He}_2^+$ is also  stable towards decay to ${\rm He}_2^+  \nrightarrow {\rm
He}^-     +  \alpha $ { with a dissociation energy of $\sim 1618$ eV.}  For the case of channel (c)
${\rm He}_2^+  \to {\rm He} + \alpha + e$, the total
energy of the sub-products
of this decay is larger  than the total energy of the ${\rm He}_2^+ $ ion in the quartet state. So,
the system is also stable towards this decay channel { with a dissociation energy of $\sim 1678$ eV.}  For the case of channel (d)
${\rm He}_2^+  \to {\rm He}^+  +  {\rm He}^+ + e$ the total energy of the
sub-products of this decay is larger  than the total energy of the ${\rm He}_2^+ $ ion in the
quartet state. { The dissociation energy in this case is $\sim 976$ eV.} For the case of channel (e)     ${\rm He}_2^+  \to {\rm He}_2^{2+} ({}^3\Pi_u) + e\,$, at $B=10000\,$a.u. the
total energy of the sub-products of this decay is larger
than the total energy of the ${\rm He}_2^+ $ ion in the quartet state, and the corresponding dissociation energy is $\sim 752$ eV.

For the case (f)     ${\rm He}_2^+  \to {\rm He}_2^{3+} (1\sigma_g) + 2 e\,$, at $B=10000\,$a.u. the
total energy of the sub-products of this decay is larger
than the total energy of the ${\rm He}_2^+ $ ion in the quartet state{ , and the corresponding dissociation energy is $\sim 1953$ eV}.

\bigskip

\begin{figure}[tb]
\begin{center}
   \includegraphics[angle=-90,width=4.5in]{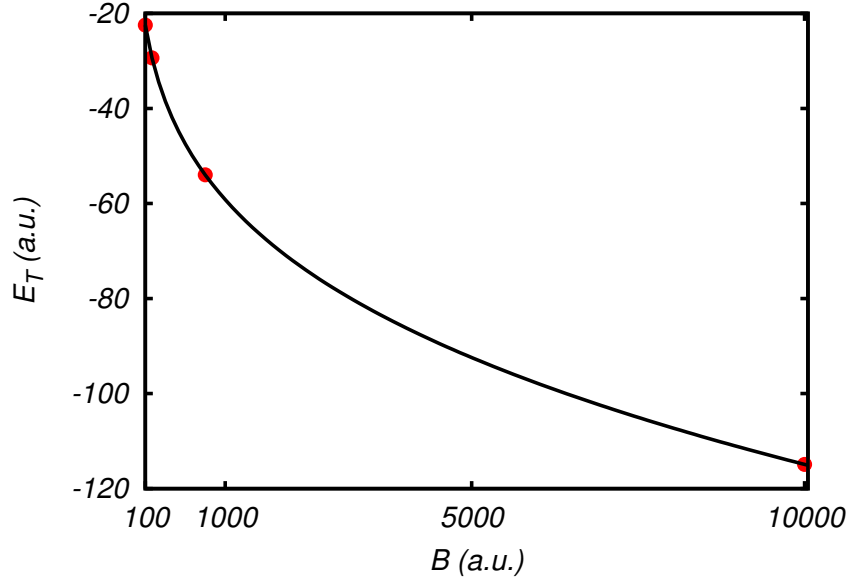}
    \caption{\label{ET}
    Total energy $E_T$ of  the ${\rm He}_2^+ $ ion in the spin-quartet state ${}^4(-3)^+ $ as a function of the magnetic field $B$, the continuous line is the fit $E_{T}(B)= -2.7582 \log^2(B) + 18.078\log(B)  - 47.4006$. }
    \end{center}
\end{figure}

\begin{figure}[tb]
\begin{center}
   \includegraphics[angle=-90,width=4in]{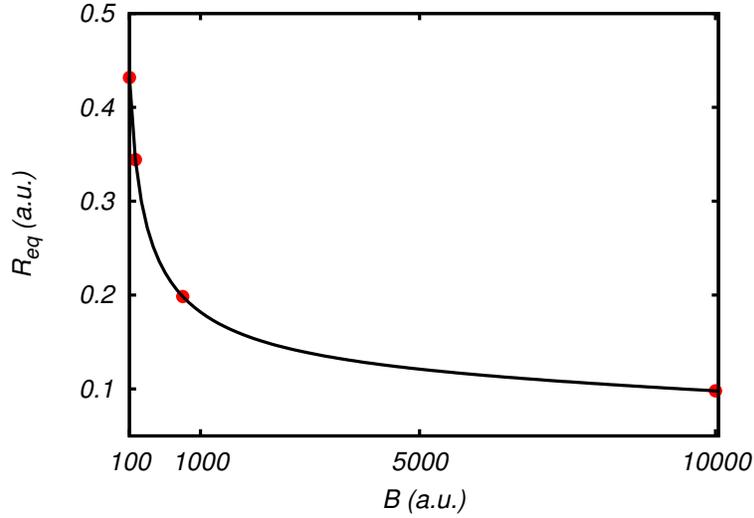}
    \caption{\label{Req}
    Equilibrium distance of the ${\rm He}_2^+ $ ion in the spin-quartet state ${}^4(-3)^+ $ as a function of the magnetic field, the continuous line is the fit
    $R_{eq}(B)=  0.01262\log^2(B)  -0.24654\log(B) + 1.29830$. }
    \end{center}
\end{figure}

\begin{figure}[tb]
\begin{center}
   \includegraphics[angle=-90,width=6in]{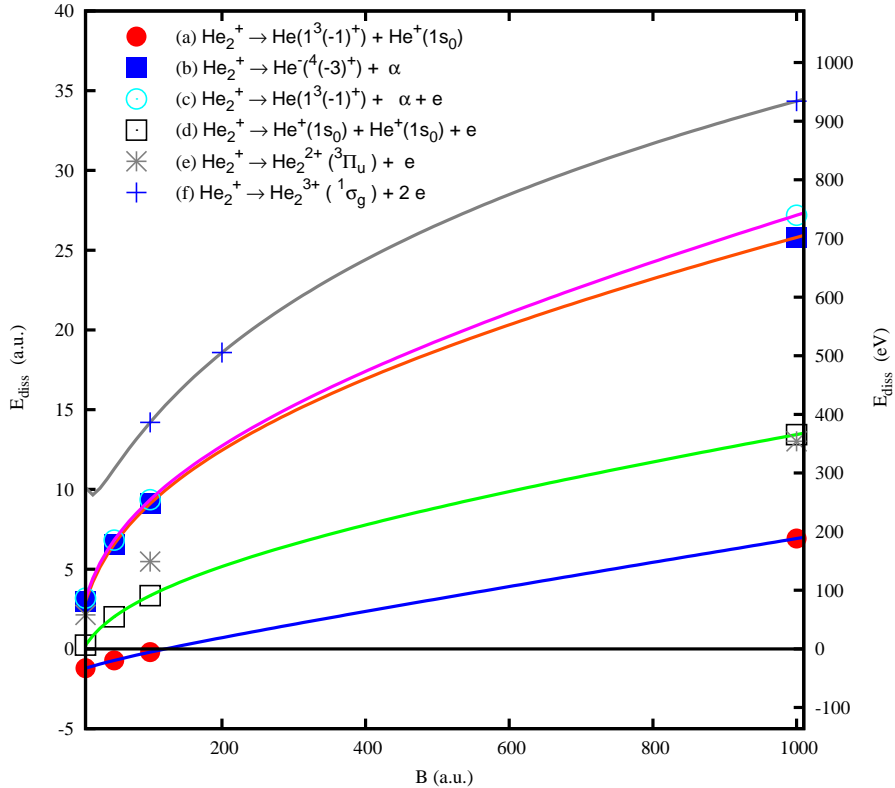}
    \caption{\label{DeltaE}
    Dissociation energies  of the ${\rm He}_2^+ $ ion in the spin-quartet state
    ${}^4(-3)^+_g $ as a function of the magnetic field { $ 10\lesssim B\leq 1000$}\, a.u. towards
    different decay channels.}
    \end{center}
\end{figure}

The total energy of  the ${\rm He}_2^+ $ ion in the quartet state ${}^4(-3)^+_g $ as a function of the magnetic field   is presented in figure~(\ref{ET}).
This figure shows that as the magnetic field is increased, the system becomes more bound. The
internuclear equilibrium distance of the ${\rm He}_2^+ $ ion in the quartet state ${}^4(-3)^+ $ as a
function of the magnetic field is presented in figure~(\ref{Req}).
This figure shows that as the magnetic field is increased, the system becomes more compact.
We have plotted in Figure (\ref{DeltaE}) { the dissociation energies, ({\it i.e.}}  difference of total energies between the state
${}^4(-3)^+_g $ of ${\rm He}_2^+$ and the energy of the final
products) for the different decay channels described in section \ref{sectiondecays}.
This plot shows that the ion ${\rm He}_2^+ $  in the
state ${}^4(-3)^+ $ becomes more bound with respect to all dissociation channels as the
magnetic field increases { and, more important, that the dissociation energies at $B\sim 1000$ a.u. lie in the window $0.1-1$ KeV which is the window of observed absorption features in the spectrum of the isolated neutron star 1E1207.4-5209 (see ~\cite{2002ApJ...574L..61S} and
\cite{2002ApJ...581.1280M} ).}

In order to have a hint about the critical magnetic field at which the state
${}^4(-3)^+_g $ of ${\rm He}_2^+$ becomes the ground state of the Coulomb system $(\alpha\alpha
e e e)$, we have collected in Table~\ref{Bcrit} a list of atomic and molecular Coulomb systems made
out of Hydrogen and/or $\alpha$-particles, as well as the Lithium atom, and the corresponding
critical magnetic fields at which the ground state is realized by a state with all spins oriented
anti-parallel to the magnetic field. A simple analysis of this table indicates that for atomic
systems (with more than one-electron (${\rm He},{\rm He}^-,{\rm Li}$)) the critical magnetic field
is rather weak being
$B_{\rm crit}\sim 0.75 - 2.21\,$a.u. For molecular type systems there seems to be two typical
ranges of values for the critical magnetic field, {\it i.e.} $B_{\rm crit}\sim 10-20\,$a.u., and
$B_{\rm crit}\sim 1000-2000\,$a.u. which is much larger than for the atomic type systems.
Perhaps, this phenomenon can be explained by the fact that electrons in a molecular system are
further apart than in the case of atoms. So, if this tendency is also valid for the case of the
${\rm He}_2^+$ molecular ion, then it is very likely that the state  ${}^4(-3)^+_g $ becomes the
ground state for magnetic fields $B\gtrsim 1000\,$a.u. Of course, a detailed study of other states
of the ${\rm He}_2^+$ molecular ion is needed in order to establish this conjecture.


\begin{figure}[tb]
\begin{center}
   \includegraphics[angle=-90,width=6in]{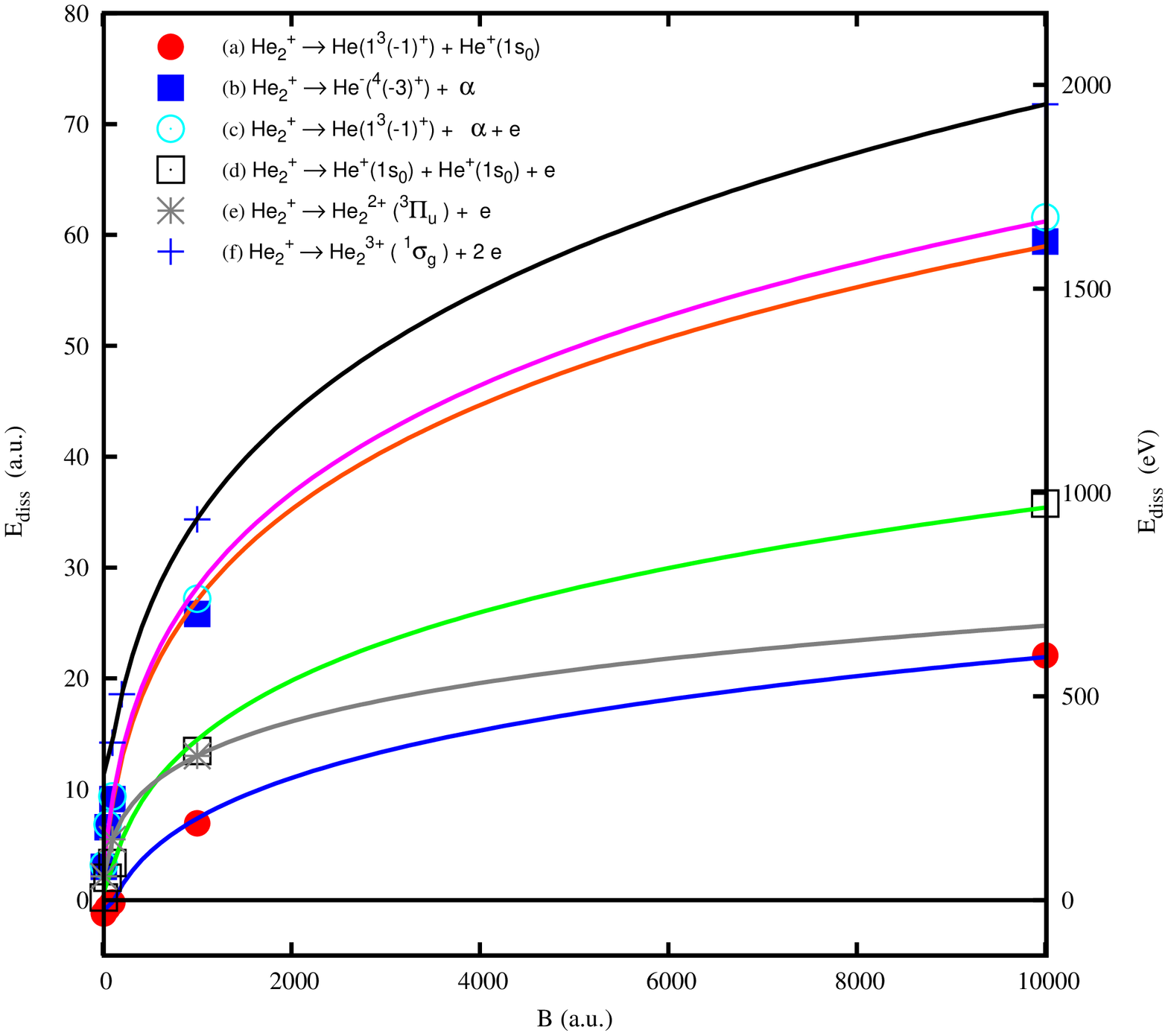}
    \caption{\label{DeltaE2}
    Dissociation energies  of the ${\rm He}_2^+ $ ion in the spin-quartet state
     ${}^4(-3)^+_g $ as a function of the magnetic field { $ 10\lesssim B\leq 10000$}\, a.u. towards different decay channels (same as Fig. \ref{DeltaE} but extended up to $B=10000$ a.u.).  }
    \end{center}
\end{figure}

\subsection{Spin $1/2$ states}
{ As for the spin $S=1/2$ states of ${\rm He}_2^+$ we have calculated the total energy of some states with $M=0,-1$ for both gerade and ungerade parities at the magnetic field $B=100$ a.u. (see Table \ref{tables12}). It is clear from such results that, for high magnetic fields $B>100$ a.u.,  these states will lay much higher in energy than the fully polarized $S=3/2$ state that we consider in our study.   For spin $S=1/2$ states the contribution to the total energy coming from the Spin-Zeeman term is $ E_{\rm Zeeman} ^{S_z=-1/2}=  - B/2$\,  a.u., while it is
$E_{\rm Zeeman}^{S_z=-3/2} = - 3/2 B$\, a.u. for the spin $S=3/2$ fully polarized state. Thus, considering only the Spin-Zeeman contribution to the total energy, the states with spin $S=1/2$ are about $\Delta E_T \sim B$\, a.u. higher that the spin $S=3/2$ states.
}

\begin{table}[thb]
\begin{center}
\begin{tabular}{|c|cc|cc|cc|}
\hline
&\multicolumn{4}{|c|}{$S=1/2$}& \multicolumn{2}{|c|}{$S=3/2$}  \\
\hline
&\multicolumn{2}{|c|}{$M=0$}& \multicolumn{2}{c|}{$M=-1$} & \multicolumn{2}{|c|}{$M=-3$} \\
&g & u & g & u & g & \\
\hline
$E\,$ (a.u.) &76.13 &  76.27 & 69.57 & 66.84 & -22.46 & \\
$R\,$ (a.u.) &1.85  &   0.79 & 0.279 & 0.79  & 0.432 & \\
\hline
\end{tabular}
\end{center}
\caption{\label{tables12}
Total energies and equilibrium distances (in a.u.) for spin $S=1/2$  states of the ${\rm He}_2^+$ ion in a magnetic field $B=100\,$a.u., in parallel configuration calculated with a trial function (\ref{ansatzs12}).
For $M=-1$ the configuration $m_1,m_2,m_3 = -1,0,0$ was used in the trial function. The energy of the spin $3/2$ state ($M=-3$ with $m_1,m_2,m_3 = 0,-1,-2$) included for comparison.}
\end{table}

 \section{Conclusions}

We have studied the stability of the  molecular Coulomb system formed by two
infinitely massive $\al$-particles and three electrons, $(\al \al e e e)$  in the range of magnetic fields { $0~\leq~B~\leq~10000\,$~a.u.},  in  a state where all electron spins are oriented antiparallel to the magnetic field direction, hence, $S_z = -3/2$.
It was further assumed that in the ground state, in order to suppress the appearance of
the Pauli force, all electrons should have different magnetic quantum numbers, in particular, if these are equal to $m_1=0,m_2=-1,m_3=-2$, the total magnetic quantum number $M=-3$. This choice looks natural physically.
The parallel configuration, for which both the molecular axis and the magnetic field direction coincide, was adopted as the optimal configuration with minimal total energy. The stability towards possible decay channels was studied variationally, using trial functions (\ref{ansatznospin}), (\ref{phi}). We found that for all studied magnetic fields $1 \leq B \leq 10000\,$~a.u., there exists a well-pronounced minimum in the potential curve of total energy {\it vs} the internuclear distance $R$ at some $R=R_{eq}$. The equilibrium distance $R_{eq}$ decreases with the magnetic field increase, hence, the system becomes more compact at large magnetic fields. At the same time the total energy is getting more negative, while the binding energy increases making the system more bound. For { $B \gtrsim 120\,$a.u.}, the ${\rm He}_2^+$ molecular ion in the state ${}^4(-3)^+_g$ becomes stable towards all possible decay channels (see Fig. (\ref{DeltaE})). In other words, the molecular system ${\rm He}_2^+$ becomes the most bound Helium specie with three electrons.
It also hints to the possible relevance of other Helium chains like ${\rm He}_2$, ${\rm He}_3$, or even the hybrid molecules like the neutral ${\rm HeH}$,  for the chemistry in a strong magnetic field.

Studying the evolution of the ground state with the magnetic field change in different Coulomb systems, with different number of electrons, we found that there is always a specific, well-defined state of maximal total spin projection, which becomes the ground state at large magnetic fields. For one-electron systems this state has $S_z = -1/2$ and $M=m_1=0$. For two-electron systems it has $S_z = -1$ and $M=-1$ with $m_1=0, m_2=-1$, see Table II. For two studied three-electron systems, Li and He$^-$, it was the state $S_z = -3/2$ and $M=-3$, \cite{2013PRL111}. It seems natural to assume that this state will be the ground state for ${\rm He}_2^+$ for a certain {\it critical} magnetic field, see Section III.
In order to find this critical magnetic field it is necessary to explore other states of ${\rm He}_2^+$, in particular, spin-quartet gerade and ungerade states with different total electron angular momentum ${}^4(0,-1,-2,-4)^+_{g,u}$. This will be done elsewhere.

Concluding we have to state that at $B=1000\,$a.u., the dissociation energy with respect to the main decay channel ${\rm He}_2^+ ({}^4(-3)^+) \to {\rm He}(1{}^3 (-1)^+)  + {\rm He}^+(1s_0) $ reaches { $E_{\rm diss} \simeq 6.54\,$a.u. ($\sim 178\,$eV)}; while the dissociation energy for the decay channel (b) into ${\rm He}^- + \al$ is { $\sim 25.8\, {\rm a.u.}  \simeq 702 \,$eV} and the  dissociation energy for the decay channel (c) into ${\rm He} + \al + e $ is { $\sim 26.8\, {\rm a.u.} \simeq 729 \,$eV}. Thus, the two latter energies are in the energy window for one of the absorption features observed for the isolated neutron star 1E1207.4-5209.

We found that in a strong magnetic field $B\gtrsim 120\,$a.u., the molecular ion ${\rm He}_2^+$ is the most bound system among the  atomic and molecular systems  containing helium and up to three electrons. Thus, this molecule may  play a particularly important role in the description of atmosphere of strongly magnetized neutron stars as it was hinted in  \cite{Kerkwijk:2007}.

\begin{acknowledgments}
  This work was supported in part by the PAPIIT grant {\bf IN108815} and CONACyT grant {\bf 166189}~(Mexico). The authors thank D. Turbiner for his valuable help in designing and optimizing the parallel programming codes in C\texttt{++}.  Also J.C.L.V. thanks PASPA grant (UNAM, Mexico) and the Centre de Recherche Math\'ematiques, Universit\'e de Montr\'eal for the kind hospitality  while on sabbatical leave during which this work was ultimately completed.
\end{acknowledgments}

\newpage

\begin{table}[h]
    \centering
\begin{center}
\begin{tabular}{|c|c|c|c|c|c|c|c|}\hline
$B$         &  \multicolumn{2}{|c|}{
${\rm He}_2^{+} ({}^4(-3)^+_g)$}   & ${\rm He}^-$           & ${\rm He}^+$&
${\rm He}_2^{3+}$& ${\rm He}_2^{2+}$& ${\rm He}$
       \\
 \cline{2-3}
  (a.u.)      &   $E_T$             &     $R_{\rm eq}$                    & $E_T({}^4(-3)^+)$
&
$E_T (1s_0)$   & $E_T(1\sigma_g)$  & $E_T({}^3\Pi_u)$&  $E_T(1{}^3(-1)^+)$  \\ \hline
 1.        &  -4.02  &  2.18   & -3.03      & -2.4410  & -                 &  -3.3745    & -2.9655 \\
 100.    & -22.46 & 0.432  & {-13.38}  & -9.5605  &  -8.2581     & -16.9917   & -13.1048 \\
1000.   & -53.98 & 0.196  & -28.18    & -20.2707 &  -19.6338  & -40.2462   & -27.1738 \\
10000. & -114.9 & 0.098  & {-55.41}  & -39.5107 & -43.1165   &  -87.255    & -53.2011 \\
\hline
\end{tabular}
\caption{\label{TableEs}
Variational results for the total energy and equilibrium distance of the {\em spin-quartet} state ${}^4(-3)^+_g$ with total angular momentum projection $M=-3$ of the He${}_2^+$ molecular ion in a magnetic field and comparison with the total energy of different subsystems.
For Helium atom ground state energy $E_T(1{}^3(-1)^+)$
Ref.~\cite{Becken:1999} for $B\lesssim 100$ a.u., while for $B\ge 1000$ a.u., the Ansatz (\ref{psiHe}) was used (see section \ref{sectionHe}). Energies for ${\rm He}^{-}$ in the state ${}^4(-3)^+$ at $B=100-10000\,$a.u. were calculated for the present study using the trial function (\ref{GStrialfunct})  (see section \ref{sectionHeminus}), while for $B=1\,$a.u. it was taken from \cite{2013PRL111}.
 For ${\rm He}^+$ the energies from the scaling relation (\ref{Hscaling}) with use of data
from \cite{Kravchenko:1996} (with electron spin Zeeman contribution included).
For ${\rm He}_2^{2+}$ energies from \cite{2006PhRvA..74f3419T}, and for ${\rm He}_2^{3+}$ results from \cite{Turbiner2006309} with the (single) electron spin Zeeman contribution added.
All molecular systems assumed in parallel configuration as optimal. All energies in a.u.}
\end{center}
\end{table}

  \begin{table}[h]
    \centering
\begin{center}
\begin{tabular}{|l|c|c|c|c|}\hline
System &   Ground State    & $M$ & $S_z$    & $B_{\rm crit}$ (a.u.)  \\ \hline\hline
 ${\rm He}^+$       &  $1s_0$        &  0 & -1/2  &  $\geq 0$  \\
 ${\rm He}_2^{3+}$  &  $1\sigma_g$   &  0 & -1/2  &  $\geq 10$  \\
 ${\rm He}$         &  $1{}^3(-1)^+$ & -1 & -1    &  $\gtrsim 0.75\ $  \\
 ${\rm H}_2$        &  ${}^3\Pi_u$   & -1 & -1    &  $\gtrsim 12.3\ $  \\
 ${\rm He}_2^{2+}$  &  ${}^3\Pi_u$   & -1 & -1    &  $\gtrsim 1100\ $  \\
 ${\rm H}_3^{+}$    &  ${}^3\Pi_u$   & -1 & -1    &  $\gtrsim 20\ $  \\
 ${\rm H}_4^{2+}$   &  ${}^3\Pi_u$   & -1 & -1    &  $\gtrsim 2000\ $  \\
 ${\rm HeH}^{+}$    &  ${}^3\Pi_u$   & -1 & -1    &  $\gtrsim 15\ $  \\
 ${\rm He}_3^{4+}$  &  ${}^3\Pi_u$   & -1 & -1    &  $\gtrsim 1000\ $  \\
 ${\rm He}^{-}$     &  ${}^4(-3)^+$  & -3 & -3/2  &  $\gtrsim 0.74\ $  \\
 ${\rm Li}$         &  ${}^4(-3)^+$  & -3 & -3/2  &  $\gtrsim 2.21\ $ \\
 \hline
\end{tabular}\caption{\label{Bcrit}
Critical magnetic field for different Coulomb systems for which the ground state
becomes a state with maximal electronic spin: all electron spins are antiparallel to the magnetic field. Results taken from\cite{2013PRL111,PhysRevA.81.042503,Becken:2000,Turbiner2006309,2006PhRvA..74f3419T}}
\end{center}
\end{table}

 \clearpage

\appendix

\section{Variational Parameters for ${\rm He}_2^+$}
\label{appendixA}

\begin{table}[hb]
\begin{center}
\begin{tabular}{| c | c c c|}
\hline
             &   \multicolumn{3}{c|}{Magnetic field $B$ in a.u.} \\
parameter    &         100   &    1000   &   10000 \\  \hline
$\alpha_{1, {\sf A}}$              &  0.93175    &    1.975   &   3.59        \\
$\alpha_{1, {\sf B}}$             &  3.34641    &    3.9     &   4.28        \\
$\alpha_{2, {\sf A}}$             &  1.41717    &    1.9     &   2.5         \\
$\alpha_{2, {\sf B}}$            &  2.0076     &    3.059   &   4.36        \\
$\alpha_{3, {\sf A}}$             &  1.53986    &    2.09    &   2.32        \\
$\alpha_{3, {\sf B}}$             &  1.03391    &    1.82    &   3.1         \\
$\alpha_{12}$             &  0.58803    &    0.46    &   0.38        \\
$\alpha_{13} $           &  0.23269    &    0.16    &   0.08        \\
$\alpha_{23}$              &  0.17702    &    0.15    &   0.2         \\
$\beta_1$                &   0.81948   &     0.93   &    0.98        \\
$\beta_2$              &   0.87464   &     0.94   &    0.983       \\
$\beta_3$               &   0.90583   &     0.957  &    0.989       \\
\hline
$R_{\rm eq}$  (a.u.)        & 0.43177    & 0.1985     & 0.098  \\
$E_T$        (a.u.)             &-22.460     & -53.978    & -114.908 \\
\hline
\end{tabular}
\caption{\label{he2pparams}${\rm He}_2^+$ in a strong magnetic field. State ${}^4(-3)^+_g$  variational parameters for the trial function (\ref{phi}).
For the evaluation of the variational total energy, 500 millions was the maximal number of points used for the numerical integration in each subdomain in the manual partitioning.
Triple cylindrical coordinates were used with 2 subdomains in each of the three $\rho$ coordinates and 7 subdomains in each of the three $z$ coordinates. The integration routine {\it cubature} \cite{GenzMalik1980} was used.}
\end{center}
\end{table}

\clearpage

\section{Variational Parameters for ${\rm He}^-$}
\label{appendixB}

\begin{table}[hb]
\begin{center}
\begin{tabular}{| c | c c c|}
\hline
                     &   \multicolumn{3}{c|}{Magnetic field $B$ in a.u.} \\
parameter            &       100   &    1000   &   10000 \\  \hline
$\alpha_{1}$         &  2.86366    & 4.30183   &   5.32027    \\
$\alpha_{2}$         &  1.91669    & 3.09290   &   3.79174    \\
$\alpha_{3}$         &  1.20906    & 1.67689   &   2.48690    \\
$\alpha_{12}$        & -0.13497    & 0.06847   &   0.37195    \\
$\alpha_{13} $       & -0.12987    & 0.02558   &  -0.04966    \\
$\alpha_{23}$        & -0.11967    & 0.05634   &  -0.08924    \\
$\beta_1$            &  0.84414    & 0.92115   &   0.97709    \\
$\beta_2$            &  0.91848    & 0.95373   &   0.99092    \\
$\beta_3$            &  0.96952    & 0.98367   &   0.99470    \\
\hline
$E$                  & -13.3772    & -28.1756  &  -55.4053    \\
\hline
\end{tabular}
\caption{\label{hemparams}${\rm He}^-$ in a strong magnetic field. State ${}^4(-3)^+_g$  variational parameters for the trial function (\ref{He-phi}).
For the evaluation of the variational total energy, 500 millions was the maximal number of points used for the  numerical integration in each subdomain in the manual partitioning.
Triple cylindrical coordinates were used with 2 subdomains in each of the three $\rho$ coordinates and 5 subdomains in two $z$ coordinates and 3 subdomains in the third $z$ coordinate which integrated over the half line. The integration routine {\it cubature} \cite{GenzMalik1980} was used.}
\end{center}
\end{table}

\clearpage

\section{Variational Parameters for ${\rm He}$ atom in spin triplet state ${1}^3(-1)^+$}
\label{appendixC}

\begin{table}[hb]
\begin{center}
\begin{tabular}{| c | c c c|}
\hline
              &   \multicolumn{3}{c|}{Magnetic field $B$ in a.u.} \\
parameter       &       100        &    1000         &   10000 \\  \hline
$\alpha_{1}$   &  2.107315    &    3.085664  &  4.589369  \\
$\alpha_{2}$   &  2.952063    &    4.151930  &  5.887318  \\
$\alpha_{12}$ &  0.145703    &    0.117778  &  0.227942  \\
$\beta_1$       &  0.895958    &    0.958737  &  0.986090  \\
$\beta_2$       &  0.825659    &    0.930747  &  0.977742  \\
\hline
$E$                 & -12.8215     &   -27.1738   & -53.2011  \\
\hline
\end{tabular}
\caption{\label{heparams}${\rm He}$ atom in a strong magnetic field. Spin triplet  state ${1}^3(-1)^+$  variational parameters for the trial function (\ref{psiHe}). }
\end{center}
\end{table}

\newpage

\bibliography{refsMagnetic}

\end{document}